\documentclass[preprint,prd,showkeys,showpacs]{revtex4}
\pdfoutput=1
\usepackage{amsfonts,amsmath,amssymb}
\usepackage{enumerate}
\usepackage{hyperref}
\usepackage{graphicx}
\newcommand{\be}{\begin{equation}}
\newcommand{\ee}{\end{equation}}

\newcommand{\bea}{\begin{eqnarray}}
\newcommand{\eea}{\end{eqnarray}}

\newcommand{\bes}{\begin{subequations}}
\newcommand{\ees}{\end{subequations}}

\newcommand{\nn}{\nonumber}

\newcommand{\cN}{{\cal N}}

\newcommand{\Tr}{\mbox{Tr}}
\newcommand{\tr}{\mbox{tr}}

\newcommand{\cp}{\mathbb {CP}}

\newcommand{\z}{\zeta}
\newcommand{\ep}{\epsilon}

\begin{document}

\title{Supersymmetric Wilson loops with general contours \\in ABJM theory }
\author{Nakwoo Kim}
\email{nkim@khu.ac.kr}
\affiliation{Department of Physics
and Research Institute of Basic Science, \\ Kyung Hee University,
 1 Hoegi-dong, Dongdaemun-gu,
 Seoul 130-701, Korea\\}



\begin{abstract}
We consider general supersymmetric Wilson loops in ABJM model, i.e. Chern-Simons-matter theory in 2+1 dimensions with $\cN=6$ supersymmetry. They are so-called Zarembo-type: the Wilson loops of our
interest have generic contours in spacetime, but the scalar field coupling is arranged 
accordingly so that
there are unbroken supersymmetries. Based on the supermatrix formulation of Wilson loops by Drukker 
and Trancanelli, we construct explicitly 1/6-BPS Wilson loops and check that their expectation value
is protected using perturbation up to two loops. We also study the dual string configuration in 
$AdS_4\times \mathbb{CP}^3$ background and check the supersymmetry. 
\end{abstract}


\maketitle

\section{Introduction}
Wilson loops are essential objects in the study of
AdS/CFT correspondence \cite{Maldacena:1997re}. 
On the gauge theory side
they provide gauge singlet operators whose vacuum expectation value characterizes the 
confining/deconfining phase. On the string theory side they are simply fundamental 
strings, and their classical solution is given as  minimal surface \cite{Maldacena:1998im}\cite{Rey:1998ik}. 
The 
importance of Wilson loops in AdS/CFT partly stems from the fact that they are amenable to both perturbative gauge theory computation \cite{Erickson:2000af} and 
a string theory computation in the region of large coupling constant \cite{Maldacena:1998im}\cite{Rey:1998ik}. It is also possible to calculate 
the expectation value of supersymmetric Wilson loops exactly using the technique of 
localization \cite{Pestun:2007rz}.

In this paper we are interested in supersymmetric Wilson loops in $\cN=6$ Chern-Simons-matter theory in $2+1$-dimensions. This theory was proposed as the gauge theory on 
M2-branes put on orbifold singularity $\mathbb{R}^8/\mathbb{Z}_k$ by 
Aharony, Bergman, Jafferis, and Maldacena (ABJM)  \cite{Aharony:2008ug}. See also the Aharony-Bergman-Jafferis (ABJ) model \cite{Aharony:2008gk} of fractional M2-branes which have different gauge group ranks, i.e. $SU(N)\times SU(M)$. In the large coupling 
region the relevant gravity dual background is $AdS_4\times \mathbb{CP}^3$ in IIA
string theory, and the Wilson loop variables are expected to be dual to fundamental strings.
Initially there was a puzzle over the matching of gravity side and gauge theory side
description for Wilson loops: On gravity side a simple straight fundamental string is 1/2-BPS, 
while constructing 1/2-BPS Wilson line in ABJM turned out to be nontrivial \cite{Drukker:2008zx}\cite{Chen:2008bp}\cite{Rey:2008bh}. 

This enigma was resolved by Drukker and Trancanelli  \cite{Drukker:2009hy} when they constructed explicitly 1/2-BPS Wilson 
loops in ABJM model using supermatrix. In contrast to the supersymmetric Wilson loops in $\cN=4$ super Yang-Mills theory, the Wilson loops in  \cite{Drukker:2009hy} have couplings to fermion fields in bi-fundamental 
representation of the gauge group $U(N)\times U(N)$. The authors of  \cite{Drukker:2009hy} considered Wilson lines which are 
straight lines  or circular loops. They are all 1/2-BPS, and are related through conformal transformation. 

Our aim here is to provide a nontrivial test of AdS/CFT for ABJM model, by considering
supersymmetric Wilson loops with generic contours. Such objects
in $\cN=4$ super Yang-Mills theory  were first constructed by Zarembo \cite{Zarembo:2002an}. One important
feature of Zarembo-type Wilson loop is that the scalar field coupling is coordinated with the spacetime contour
so that there is unbroken supersymmetry. If the contour is straight line it is 1/2-BPS, while if the loop is embedded in $\mathbb{R}^2$ ($\mathbb{R}^3$) it is 1/4-BPS (1/8-BPS).

The study of less-than-half BPS Wilson loops in 
ABJM theory was initiated in \cite{Griguolo:2012iq}. Cusp-like, or piecewise linear Wilson lines are studied 
and it was explicitly checked that for planar ones the supersymmetry is broken to 1/6.  
Then the extension to 
Zarembo-type Wilson loops of arbitrary shape was presented in \cite{Cardinali:2012ru}. 

  In this paper we aim to  elaborate upon the 
result in \cite{Cardinali:2012ru}. In particular, we construct a Zarembo-type circular Wilson loop explicitly and 
check if its expectation value vanishes perturbatively, up to two loops. We also demonstrate the supersymmetry of the
dual string configuration, using the extremal surface satisfying the pseudo-holomorphicity condition in \cite{Dymarsky:2006ve}. 

We note that recent papers which study supersymmetric Wilson lines in ABJM model include
\cite{Kluson:2008wn,Kluson:2009vq,Henn:2010ps,Lee:2010hk,Bianchi:2011rn,Wiegandt:2011uu,
Kim:2012tu,Klemm:2012ii,Marmiroli:2012ny,Bianchi:2013zda,Bianchi:2013pva}.
For conventions and the Feynmann rules for perturbative computations we use in this paper, the readers are 
referred to  \cite{Griguolo:2012iq}.

This paper is organized as follows. In Sec.\ref{s2} we introduce the supermatrix Wilson 
loop formalism \cite{Drukker:2009hy} and discuss the role of boundary condition for supersymmetry condition. 
In Sec.\ref{s3} we present the configuration of 1/6-BPS Wilson loops
explicitly. Then in Sec.\ref{s4} we describe the perturbative computation of our Wilson loops
and check it vanishes up to two-loops. In Sec.\ref{s5} we provide the dual picture and show 
the pseudo-holomorphic string solution is 1/6-BPS using $\kappa$-symmetry projection. 
We then conclude in Sec.\ref{s6}.

\section{Supersymmetric Wilson loops in ABJM theory}
\subsection{The supermatrix formalism and the boundary condition of Wilson loops}
\label{s2}
Let us first recall the supermatrix Wilson loop of Drukker and Trancanelli \cite{Drukker:2009hy}. The Wilson loop operator is given as a path-ordered exponential integral defined by a curve $x^\mu (\tau)$ in spacetime, 
\be
W = {\rm Tr} \, {\cal P} \exp \left( \int L d\tau \right) . 
\ee
Here the exponentiated object $L$ is a supermatrix,
\be
L = 
\begin{pmatrix}
i  A_\mu \dot{x}^\mu + \frac{2\pi}{k} |\dot x | M^{I}_{\;J} C_I \bar{C}^J & \sqrt{\frac{2\pi}{k}}|\dot x| \eta^\alpha_I \bar{\psi}^I_\alpha  \\ 
 \sqrt{\frac{2\pi}{k}}|\dot x| \psi_I^\alpha \bar{\eta}_\alpha^I   &  i \hat{A}_\mu \dot{x}^\mu + \frac{2\pi}{k} |\dot x | \hat{M}_{I}^{\;J} \bar{C}^J C_I
 \end{pmatrix} 
 \, . 
\label{sm}
\ee
We note that in this article the Wilson lines are defined along a space-like curve. 
When necessary, we will consider Euclidean $\mathbb{R}^3$ as the spacetime.  The choice above is made to be consistent with this convention. As well-known, the ABJM model has gauge group
$U(N)\times U(N)$. More generally the Aharony-Bergman-Jafferis (ABJ) model  \cite{Aharony:2008gk}, a Chern-Simons model which incorporates fractional M2-braneABJ model may have different ranks for the gauge group,
i.e. $U(N)\times U(M)$. The above supermatrix $L$ is in general $(N+M)\times (N+M)$, and
the top-left block is $N\times N$, etc. $k$ is the Chern-Simons coefficient. 

In the above $A_\mu, \hat{A}_\mu$ represent the two gauge fields. $C_I, I=1,2,3,4$ 
are the bifundamental scalar fields and $N\times M$. $\bar{C}^I$ are their conjugate
fields. The same applies to fermionic fields $\psi_I, \bar{\psi}^I$. On the other hand,
$x,M,\hat{M}, \eta, \bar{\eta}$ are $\tau$-dependent parameters, defining the Wilson line. In particular, they encode the shape of the Wilson 
loop in spacetime ($x^\mu$) or in the internal space $\cp^3$ ($M$ and $\hat{M}$).
Also note that the spinorial parameters $\eta,\bar{\eta}$ are commuting variables. Thus in the 
end the diagonal blocks are commuting, while the off-diagonal blocks are anti-commuting quantities.

Before we embark on the 1/6-BPS Wilson lines, let us make a comment on the relation between 
the boundary condition of the fields along the Wilson line and the choice of taking either
ordinary trace or supertrace.
For convenience let us record a few terms in the
path-ordered exponent. 
\be
W = \Tr \,\mathbb{I} + \int^{\infty}_{-\infty} d\tau\, {\rm Tr} L  + 
\int^{\infty}_{-\infty} d\tau_1  \int^{\tau_1}_{-\infty} d\tau_2 \, {\rm Tr} \, 
L(\tau_1) L(\tau_2) + \cdots
\label{wexp}
\ee
When one considers the supersymmetric variation of $W$ for straight line with constant 
$M,\hat M, \eta, \bar \eta$ chosen appropriately, schematically the result is  (see eq.(2.12) of \cite{Drukker:2009hy} with  $\eta\bar\eta=2i$)
\be
\delta W \propto \Tr \left[ \int^\infty_{-\infty} d\tau 
\begin{pmatrix}
C \psi & \cr & \psi C
\end{pmatrix} 
-
\int^\infty_{-\infty} d\tau_1 
\int^\infty_{\tau_1} d\tau_2 
\begin{pmatrix}
\partial_{\tau_1} C (\tau_1) \psi (\tau_2) & 
\cr
 & -  \psi (\tau_1) \partial_{\tau_2} C (\tau_2) 
\end{pmatrix}
\right]
\ee
Integrating the second term once, it is obvious that the bulk contribution cancels the first 
term for both top-left and bottom-right blocks. The boundary term of this integration was
simply ignored in \cite{Drukker:2009hy}, but we would like to demand that they should 
cancel out eventually as well. It is easy to check that
\be
\delta W \propto \Tr \int^\infty_{-\infty}  d\tau
\begin{pmatrix}
C(-\infty) \psi(\tau) & \cr
& \psi(\tau) C(\infty)
\end{pmatrix}
\ee
If we impose the periodic boundary condition $C(-\infty)=C(\infty)$, which is obviously a natural choice for 
bosonic fields $C^I$, the variation is zero when we take
the supertrace. In this paper we are interested in generic contour generalization of Drukker-Trancanelli 1/2-BPS Wilson lines with trivial expectation values. We may discretize such loops
as a composite of infinitesimal line elements, and clearly taking supertrace one can preserve 
some supersymmetry with this convention. We discuss the remaining supersymmetry in more detail in the next sections.

\subsection{1/6-BPS Wilson loop with  generic contours}
\label{s3}
\subsubsection{Argument using cusp Wilson lines}
Using the supermatrix representation above, Drukker and Trancanelli \cite{Drukker:2009hy} showed that
for a specific choice of $x^\mu, M^I_{\,\,J}$ the Wilson loop operator can be
1/2-BPS. For the class of BPS operators with unbroken Poincar\'e supersymmetry, the 
Wilson line is defined along 
a straight line, e.g. $x^\mu = (0,0,\tau)$. The choice of $M,\hat{M}$ is made
via a point in $\cp^3$. In other words, 
\be
M^I_{\; J} = \hat{M}^I_{\; J} = - 2 \frac{\bar{z}^I z_J}{|z|^2} + \delta^I_J \, . 
\label{mm}
\ee
Then one can show that if we choose $z_I$ to be constant, e.g. $z=(1,0,0,0)$ and
$M=\hat{M}={\rm diag} (-1,1,1,1)$ the
Wilson loop preserves 1/2 supersymmetry. The value of spinorial parameters 
$\eta, \bar{\eta}$ is also crucial, and as a $d=3$ spinor they are eigenspinors of
$\dot{x}^\mu \gamma_\mu$ with normalization $\eta\bar{\eta}=2i$. Note that, just
as $M$ and $\hat{M}$ are in principle independent of each other, $\eta$ and $\bar\eta$ are independent
and do not need to be conjugate to each other.

The ABJM model has ${\cal N}=6$ supersymmetry, and the transformation parameters
can be written as $\theta^{IJ}_\alpha$. Here $I,J$ are $SU(4)$ indices and antisymmetrized,
and $\alpha=1,2$ are Dirac indices in $d=3$. For the particular 1/2-BPS Wilson loop operators above, 
the remaining supersymmetries are shown to be generated by the following set of 
parameters \cite{Drukker:2009hy}. 
\be
\bar{\theta}^{12}_+, \bar{\theta}^{13}_+, \bar{\theta}^{14}_+, \bar{\theta}^{23}_-, \bar{\theta}^{34}_-, 
 \bar{\theta}^{42}_-
\, . 
\ee
The rule for preserved supersymmetry is very simple: for $\bar\theta^{1i}\dot{x}^\mu \gamma_\mu=
\bar\theta^{1i}, \quad \bar\theta^{jk}\dot{x}^\mu\gamma_\mu = - \bar\theta^{jk}, 
i,j,k=2,3,4$ when
$z^I = \delta^I_1$. 

Now let us rationalize the existence of less supersymmetric Wilson loop
operators, using a simple configuration.  
Consider another  Wilson loop which is directed opposite to the one above in spacetime ($x^\mu=(0,0,-\tau)$)
and on a different point
in $\cp^3$,  $z=(0,1,0,0)$.
Then obviously the supersymmetry should be now generated by 
\be
\bar{\theta}^{21}_-, \bar{\theta}^{23}_-, \bar{\theta}^{24}_-, \bar{\theta}^{13}_+, \bar{\theta}^{34}_+, \bar{\theta}^{41}_+, 
\ee
If we look at the intersection of these two sets, we realize it has four elements
\be
 \bar{\theta}^{13}_+, \bar{\theta}^{14}_+,\bar{\theta}^{23}_-, \bar{\theta}^{24}_- . 
\ee
In other words, the composite of these two Wilson lines preserve 4 out of 12 original Poincar\'e
supersymmetry, i.e. 1/3-BPS. 

Of course one can glue them together as two half-lines, and obtain a Wilson loop with a cusp of
deflection angle $\pi$. Although at first sight it looks like we have $\pi/2$ rotation with $\z^I$, 
we should recall that they are in spinor representation of $SO(6)$. 
$z=(1,0,0,0)$ into $z=(0,1,0,0)$ corresponds to $\pi$ rotation in the $SU(2)$ subsector made of
$z_1,z_2$, as multiplication of $\exp (-i  \pi \sigma_2/2 )$.

In fact the Wilson lines which are piecewise linear with cusps are studied recently in \cite{Griguolo:2012iq}. In those works Wilson lines on a plane are considered and it is checked that
for a generic deflection angle the configuration is 1/6-BPS. For the supersymmetry of 
Wilson loops with generic contour $x(\tau)$, we can consider discretizing them into a 
collection of infinitesimal linear intervals. When the deflection angle, or the curvature in spacetime and in internal $\cp^3$ are synchronized as above, we expect the same 1/6 supersymmetry 
is preserved by the Wilson loop. One can see the mechanism in the following way. Originally
the supersymmetry parameters (and also the supercharges) constitute $(6,2)$ representation of
$SU(4)_R\times SO(3)_{Lorentz}$. If we restrict to $SU(2)_R\subset SU(4)_R$, then the preserved 
supercharges are simply the singlets of the diagonal $SU(2) \subset SU(2)_R\times SO(3)_{Lorentz}$.

The choice of parameters $M(\tau),\hat{M}(\tau),\eta(\tau),\bar\eta(\tau)$ for arbitrarily shaped supersymmetric Wilson loop with $x(\tau)$ is
reported in  \cite{Cardinali:2012ru}.
Instead of verifying the formulas of \cite{Cardinali:2012ru}, in this paper we will establish 
evidences for supersymmetry by showing that the vacuum expectation value of such 1/6-BPS
Wilson loops are trivial. In the forthcoming sections we construct explicitly a general class of such 
Wilson loops, and compute their expectation values perturbatively to check it is protected.

\subsubsection{General 1/6-BPS Wilson loops}
As stated above, the idea is to relate $SU(2)\in SU(4)$ global R-symmetry of ABJM theory
with the Lorentz group, and consider singlet of diagonal $SU(2)$. In particular, we allow the Wilson loop to take a generic contour in Euclidean spacetime $\mathbb{R}^3$. So $x^\mu(\tau)$ is an arbitrary 3-vector at each $\tau$. We choose to denote it via $\tau$-dependent $SO(3)$ rotation from 
the BPS Wilson line of \cite{Drukker:2009hy}, in terms of Euler angles
$\theta(\tau),\phi(\tau)$.
\be
v(\tau) \equiv 
\frac{\dot{x}(\tau)}{|\dot{x}(\tau)|} = R_{31}(\theta) R_{12}(\phi) \begin{pmatrix} 0 \\0\\ 1 \end{pmatrix} = \begin{pmatrix} \sin\theta\cos\phi \\ \sin\theta \sin\phi \\ \cos\theta \end{pmatrix}
\ee
Also we are to employ the same rotation for $z_I,\eta^\alpha$. In particular, 
\be
\bar{\eta} (\tau) = e^{-i \phi \sigma_3/2} e^{-i \theta\sigma_2/2}\bar{\eta}(0) . 
\ee
Here $\eta(0) = (1 , 0)^T$, and 
\be
\eta^\alpha(\tau) = \sqrt{2}
\begin{pmatrix} e^{+i\phi/2 } \cos \frac{\theta}{2} \\ e^{-i\phi/2 }\sin \frac{\theta}{2} 
\end{pmatrix}
, \quad
\bar{\eta}_\alpha(\tau) =  \sqrt{2}i
\begin{pmatrix} e^{-i\phi/2 } \cos \frac{\theta}{2} \\ e^{+i\phi/2 }\sin \frac{\theta}{2} 
\end{pmatrix} . 
\ee
We rotate $z_I \,(I=1,2)$ in the same way as $\bar{\eta}_\alpha(\tau)$, i.e.
\be
z_I =\bar{z}^{I*}=\begin{pmatrix} 
e^{-i\phi/2 } \cos \frac{\theta}{2} & e^{+i\phi/2 }\sin \frac{\theta}{2} & 0 & 0 
\end{pmatrix}^T 
\ee
$M,\hat M$ are still defined according to \eqref{mm}, and $\hat M = M^T=M^*$.
 
Let us here list a few useful properties of the parameters $M, \eta$  chosen as above:
\bea
\eta^\alpha (\tau) \bar{\eta}_\alpha (\tau) &=& 2i ,
\\
(\eta(v_2) \gamma^\mu\bar{\eta}(v_1))(\eta(v_1)\bar{\eta}(v_2))
&=& -2 ( v^\mu_1 + v_2^\mu + i \epsilon^\mu_{\,\,\,\nu\rho} v_1^\nu v_2^\rho )  ,
\\
\tr M(\tau_1)M(\tau_2) = \tr \hat M(\tau_1) \hat M(\tau_2) &=& 2 (1+ v_1 \cdot v_2 ) .
\label{tr2}
\eea

\section{Perturbative calculations}
\label{s4}
Now let us move to the perturbative computations. For ABJM model the supersymmetric 
Wilson loops of a different class with $\tr M=0$ were studied in \cite{Drukker:2009hy,Chen:2008bp,Rey:2008bh}. We choose to employ the conventions
and results of \cite{Griguolo:2012iq}. And
 we denote the $(n+1)$-th term in 
\eqref{wexp} by $W^{(n)}$. 

The first term simply gives the dimensionalities of gauge groups, e.g. the top-left part is 
\be
W^{(0)}_{top-left} = \Tr \langle 1 \rangle = N . 
\ee
So when we consider the entire supermatrix, 
\be
W^{(0)} = N-M . 
\ee
Because of supersymmetry this vacuum energy computation is exact to all orders. 

Next, we consider the following
\bea
W^{(1)}_{t.l.} 
&=&
  \Tr \int d\tau \langle i A_\mu \dot{x}^\mu 
+ \frac{2\pi}{k} |\dot{x}| M^I_{\; J} C_I \bar{C}^J \rangle
\nn\\
&=& 
\frac{2\pi}{k} \Tr  \int d\tau |\dot{x}|
M^I_{\; J}(\tau)
\langle
C_I(x(\tau)) \bar{C}^J (x(\tau))
\rangle
\nn\\
&=&
\frac{2\pi NM}{k}  \left(\int d\tau \, |\dot{x}|\tr  M \right) D(0) . 
\eea
Here we have denoted the scalar propagator by $D(x)$. More specifically,
\be
\langle C_I(x)_i^{\,\, \hat{j}} \bar{C}^J(y)_{\hat{k}}^{\,\, l} \rangle = \delta^J_I \delta^l_i \delta^{\hat{j}}_{\hat{k}} D(x-y) , 
\, \quad
D(x) = \frac{1}{4\pi |x|} . 
\ee
We also made use of the fact that the one-point function of gauge field is zero. And we may also use the fact that 
$\tr M (\tau) = 2 $ for the class of supersymmetric Wilson loops we are interested in.  So, up to this order, we have
\be
W^{(1)} = \frac{8\pi NM}{k} L D(0) . 
\ee
where $L$ is the length of Wilson loop. 

We then consider the next-order correction to this term. Obviously we should consider here 
the correction to the scalar field propator, but it is known to be zero for the ABJM mode due to supersymmetry. 

 Now let us move to the next order in the expansion of \eqref{wexp}. We have 
 \be
 W^{(2)}_{t.l.} 
 = \int_0^1 d\tau_1 \int^{\tau_1}_0 d\tau_2
\left(  - \langle {\cal A}(\tau_1) {\cal A}(\tau_2) \rangle
+ \frac{2\pi}{k}|\dot{x}_1||\dot{x}_2|
\langle(\eta \bar{\psi})_1 (\psi \bar{\eta})_2 \rangle
 \right) . 
 \ee 
 Here we denote as shorthand notation 
 \be
 x_1 = x(\tau_1) , \quad x_2 = x(\tau_2) \quad \mbox{etc.} 
 \ee
 We have several terms upon expansion, and for convenience we write them as follows and consider one by one. 
 \be
 W^{(2)}_{t.l.} = \sum_{n=1}^4 W^{(2)}_n , 
 \ee   
where
 \bea
 W^{(2)}_1 &=& -  \Tr \int_0^1 d\tau_1 \int^{\tau_1}_0 d\tau_2
 \, \langle
 A_\mu (x_1) A_\nu (x_2) 
 \rangle
 \dot{x}_1^\mu \dot{x}_2^\nu
 \\
 W^{(2)}_2 &=& \frac{2\pi i}{k} \Tr\int_0^1 d\tau_1 \int^{\tau_1}_0 d\tau_2
 \langle A_\mu (x_1) C_I(x_2) \bar{C}^J(x_2) \rangle
 \dot{x}_1^\mu |\dot{x}_2| M^I_{\; J}(\tau_2) + (1\leftrightarrow 2)
 \\
 W^{(2)}_3 &=&  \frac{4\pi^2}{k^2} \Tr\int_0^1 d\tau_1 \int^{\tau_1}_0 d\tau_2
 \, M^I_{\; J}(\tau_1) M^K_{\; L}(\tau_2) |\dot{x}_1||\dot{x}_2|
 \nn\\
 & & \times \langle C_I(x_1) \bar{C}^J(x_1) C_K(x_2) \bar{C}^L(x_2) \rangle 
 \\
 W^{(2)}_4 &=& \frac{2\pi}{k} \Tr\int_0^1 d\tau_1 \int^{\tau_1}_0 d\tau_2
 |\dot{x}_1| |\dot{x}_2| \eta_I^\alpha(\tau_1) \bar{\eta}(\tau_2)^J_\beta \langle \bar{\psi}^I_\alpha (x_1)\psi_J^\beta(x_2) \rangle .
 \eea
 
Now we need  the  propagator of the gauge fields. At tree level, it is given as follows
 \be
 \langle 
 (A_\mu)_i^{\; j} (x) (A_\nu)_k^{\; l} (y) 
 \rangle
 = \delta^j_k \delta^l_i \frac{i}{2k} \epsilon_{\mu\nu\rho} \partial^\rho_x
 \frac{1}{|x-y|} . 
 \label{vtree}
 \ee
 Then we have 
  \be
 W^{(2)}_{1 tree} =  \frac{iN^2}{2k}  \int_0^1 d\tau_1 \int^{\tau_1}_0 d\tau_2 \frac{\epsilon_{\mu\nu\rho}\dot{x}_1^\mu\dot{x}_2^\nu(x_1-x_2)^\rho}{|x_1-x_2|^3} . 
 \ee
 This expression is zero for a planar contour,  but in general it is not zero. This is related to the self-linking number of the contour (see e.g. \cite{Guadagnini:1989am} for further discussion), and an explicit computation requires regularization known as the framing procedure. This turns out to be cancelled by $W^{(2)}_4$, as we will explain later. 
 
 We are doing the computation up to 
 ${\cal O}(k^{-2})$ here, so we need to include the one-loop correction to \eqref{vtree}. This extra
 contribution is 
 \be
 W^{(2)}_{1oneloop} = -\frac{N^2M }{k^2} \int^1_0 d\tau_1 \int^{\tau_1}_0 d\tau_2
 \frac{\dot{x}_1 \cdot \dot{x}_2 }{|x_1-x_2|^2} 
 ,
 \ee
 up to a singular gauge transformation. For the one-loop vector field propagator in Chern-Simons-matter 
 theories, see e.g. \cite{Gaiotto:2007qi}. 
 
We now turn to the next term. For this we need to know the cubic interaction vertex for two scalars and one vector. This comes from the gauge covariantized kinetic term of scalar fields. 
It is straightforward to read off the Feynman rule for vertex interaction, and the result is 
\bea
\langle A_\mu (x) C_I(y) \bar{C}^J (z) \rangle
&=& 
\frac{i N^2M}{4\pi^2} \delta^J_I\int d^3w 
\Big(
\epsilon_{\nu\mu\rho} \frac{i}{2k} \partial^\rho_w \frac{1}{|w-x|} \frac{1}{|w-z|}
\partial^\nu_w \frac{1}{|w-y|}
\nn\\
&& - 
\epsilon_{\nu\mu\rho} \frac{i}{2k} \partial^\rho_w \frac{1}{|w-x|} \frac{1}{|w-y|}
\partial^\nu_w \frac{1}{|w-z|}
\Big) . 
\eea
Since in the Wilson loop computation  we are to take $y=z$,  it is obvious that $W^{(2)}_2=0$
\footnote{This term was argued to be zero in previous papers, but for different reasons. Rey et. al. \cite{Rey:2008bh} noticed this term in the end should involve $\tr M$ which is zero for the 1/6-BPS Wilson loop they were considering. On the other hand, Drukker, Plefka and Young \cite{Drukker:2008zx} pointed out this term vanishes for planar Wilson loop, irrespective of the value of $\tr M$.}.

We move on to the next term. Of course we consider 1PI diagrams only, and by contracting
the flavor indices we have 
\be
W^{(2)}_3 = \frac{N^2M}{4k^2} \int_0^1 d\tau_1 \int^{\tau_1}_0 d\tau_2 \,\frac{|\dot{x}_1||\dot{x}_2|\tr (M(\tau_1) M(\tau_2))}{|x_1-x_2|^2} . 
\ee  
We need to know $\tr M_1 M_2$ for supersymmetric Wilson loops, and as presented 
earlier \eqref{tr2}, we may substitute
\be
\tr M_1 M_2 = 2 \left( 1 + \frac{\dot{x}_1\cdot\dot{x}_2}{|\dot{x}_1||\dot{x}_2|}
\right) . 
\ee

We turn to the last term at second order, for which we need the fermion propagator,
\be
\langle \psi_I (x) \bar{\psi}^J (y) \rangle = \frac{i}{4\pi}\delta_I^J \gamma^\mu\partial_{x^\mu}\frac{1}{|x-y|} . 
\ee
So 
\bea
W^{(2)}_4
&=&
\frac{NM}{4k} \int_0^1 d\tau_1 \int^{\tau_1}_0 d\tau_2
 |\dot{x}_1| |\dot{x}_2| (\eta_I(\tau_1) \gamma^\mu \bar{\eta}^I(\tau_2))
 \partial_{x^\mu_1} \frac{1}{|x_1-x_2|}
 \nn\\
&=&
 \frac{NM}{2k} \int_0^1 d\tau_1 \int^{\tau_1}_0 d\tau_2
 |\dot{x}_1| |\dot{x}_2| \left(
 \frac{\dot{x}_1}{|\dot{x}_1|}+\frac{\dot{x}_2}{|\dot{x}_2|} - i \frac{\dot{x}_1 \times\dot{x}_2}{|\dot{x}_1||\dot{x}_2|}
 \right)\cdot
  \frac{(x_1-x_2)}{|x_1-x_2|^3} . 
\eea
The first two terms can be integrated once. Then one can easily check that these linearly divergent terms cancel the divergence of $W^{(1)}$ exactly. And the third term with $\dot{x}_1\times \dot{x}_2$ exactly cancels the self-linking number term $W^{(2)}_1$.

Collecting the results so far, we have
\be
W^{(1)}_{t.l.}+W^{(2)}_{t.l.} = -\frac{NM}{k} \int^1_0 d\tau \frac{|\dot{x}(\tau)|}{|x(\tau)-x(0)|}
+ \frac{N^2 M}{2k^2} \int_0^1 d\tau_1 \int^{\tau_1}_0 d\tau_2 \,\frac{|\dot{x}_1||\dot{x}_2|- \dot{x}_1\cdot\dot{x}_2}{|x_1-x_2|^2} . 
\label{wex}
\ee
This is the result for the $N\times N$ block, and for the final answer we should include the $M\times M$ block too. Upon taking the supertrace, the first term cancels and the second term is nonzero in general, but also vanishes for the ABJM model, i.e. if $N=M$. 
Recapitulating, we have established that up to ${\cal O}(1/k^2)$ the expectation value of 
our Wilson loops vanishes for the ABJM model. 

\section{Worldsheet solution and Pseudo-holomorphicity}
\label{s5}
In this section we will present the string theory side description of the generic contour 
Wilson loop we considered in the previous sections.  Let us start with the IIA supergravity solution 
$AdS_4\times \mathbb{CP}^3$, with metric
\be
ds^2_{10}= \frac{R^3}{4k}(ds^2_{AdS_4} + 4 ds^2_{\mathbb{CP}^3}) . 
\ee
We use the usual Poincare coordinates for $AdS_4$, 
\be
ds^2_{AdS_4} = r^2(-dt^2 + dx^2+dy^2) + \frac{dr^2}{r^2}
\ee
And for $\cp^3$ we take the following parametrization,
\bea
ds^2_{\cp^3} &=& \frac{1}{4} \left[ d\alpha^2 + \cos^2\frac{\alpha}{2} (d\theta^2_1 + 
\sin^2\theta_1 d\varphi^2_1) + \sin^2\frac{\alpha}{2} ( d\theta^2_2 + \sin^2\theta_2 d
\varphi^2_2)
\right.
\nn\\
&& 
\left.
+ \sin^2\frac{\alpha}{2} \cos^2\frac{\alpha}{2} (d\chi + \cos\theta_1 d\varphi_1 
-\cos\theta_2 d\varphi_2)^2 
\right] . 
\eea
Here $R$ sets the radius of curvature, and $k$ is related to the order of the orbifolding action \cite{Aharony:2008ug}.

In order to check the supersymmetry of probe strings, we need the Killing spinor and its supersymmetry projection 
rule. We will study fundamental string configuration which is confined to the subspace
$AdS_4\times S^2$, where the two-sphere is  parametrized by 
$\theta_1,\varphi_1$. For simplicity we set the rest of parameters to zero. After renaming 
$\theta_1=\theta,\varphi_1=\phi$,
the Killing spinor in $AdS_4\times S^2$ is reduced to (we follow the convention in e.g. \cite{Kim:2010ab})
\bea
\epsilon &=&
e^{\frac{\theta}{4} (\hat{\Gamma}\Gamma_5 - \Gamma_6 \Gamma_{11})} e^{\frac{\phi}{4} (\Gamma_{56} - \Gamma_{11}\hat{\Gamma})} 
 \nn\\
 & \times & 
\left\{ 1 + \frac{r}{2} ( t \Gamma^{03}(1-\Gamma^{012}) - x \Gamma^{13}(1-\Gamma^{012}) - y \Gamma^{23}(1-\Gamma^{012}) ) 
\right\}
r^{\frac{\Gamma^{012}}{2}}
\epsilon_0 .
\label{ks}
\eea

Here and below we use frame indices for gamma matrices. The ordering we use is 
$t,x,y,r,\alpha,\theta_1,\varphi_1,\theta_2,\varphi_2,\chi$. Then the prjection rule for 24 out of 32 spinors
 is given as 
\be
\left( \Gamma^{0123} + ( \Gamma^{49} + \Gamma^{56} + \Gamma^{78} ) \Gamma_{11}
\right)
\epsilon = 0 .
\label{proj}
\ee
This condition can be easily obtained from the dilatino variation of IIA supergravity. Since \eqref{proj} should be satisfied at every point of the
spacetime, one should demand it on $\epsilon_0$ as well.
The $\kappa$-symmetry prescription requires that $\Gamma_\kappa$ projection defined through the induced 
worldvolume form should be compatible with the Killing spinor of the background \eqref{ks}.
As a warm-up one may start with string stretched along $r$-direction. Obviously
$
\Gamma_{\kappa} = \Gamma^{03} 
$.
If we set for simplicity  $\theta=\phi=x=y=0$, the Killing spinor is simplified to 
\be
\epsilon = \left[ 1 + \frac{rt}{2} \Gamma^{03} (1 - \Gamma^{012} ) \right]
r^{\frac{\Gamma^{012}}{2}} 
\epsilon_0
\ee
Then one can check 
$
\Gamma^{03}\Gamma_{11} \epsilon = \pm\epsilon
$
is satisfied if we impose the same projection 
$
\Gamma^{03}\Gamma_{11}  \epsilon_0 = \pm\epsilon_0 , 
$
so this configuration is 1/2-BPS.

Now we consider a Zarembo-like solution with circular contour in both a spatial subspace of $AdS_4$ and $S^2$. We set $t=0$ and using $z,\sigma$ as worldsheet coordinates 
the solution is 
\bea
x &=& -\sqrt{\rho^2-z^2} \sin\sigma, \quad y = \sqrt{\rho^2-z^2}\cos\sigma, \quad
 r = 1/z , \quad 
\nn \\
\theta &=& \tan^{-1} (r \sqrt{\rho^2-z^2}) 
, \quad \phi = \sigma . 
\eea
On the boundary the string worldsheet is a circle with radius $\rho$. 
This configuration satisfies the string equation of motion and in particular it is 
pseudo-holomorphic in the sense of \cite{Dymarsky:2006ve}. In order to check 
that, we first note that the induced metric and K\"ahler form on the worldsheet are
\bea
ds^2_{ws} 
= 
\frac{z^2+\rho^2}{z^2(\rho^2-z^2)} dz^2 + \frac{\rho^4-z^4}{z^2\rho^2} d\sigma^2
, \quad 
j_{ws} = \frac{\rho^2+z^2}{z^2\rho} dz \wedge d\sigma .
\eea

To endow a complex structure to $AdS_4\times S^2$ it is advantageous to write the 
metric in the following way. 
\bea
ds^2 &=& \frac{R^3}{4k} \left[
r^2( -dt^2 + dx^2 + dy^2) + \frac{dr^2}{r^2} + d\theta^2 + \sin^2 \theta d\phi^2 \right]
\nn\\
&=& 
\frac{R^3}{4k} \left[
(X^2+Y^2+Z^2) (-dt^2 + dx^2 + dy^2 ) + \frac{dX^2+dY^2+dZ^2}{X^2+Y^2+Z^2}
\right] .
\eea
We introduced $\mathbb{R}^3$ coordinates $X,Y,Z$ from $r,\theta,\phi$ in the obvious
way. The complex structure in the target space of our choice is so that 
$J = dx\wedge dX + dy\wedge dY + dt \wedge dZ$.  It is then straightforward to check 
 the pseudo-holomorphicity condition $j_{ws} \cdot P \cdot J = P$ with 
$P_\alpha^M = \partial_\alpha X^M$.

Now we can  compute $\Gamma_\kappa$, using the induced volume form on the worldsheet.  
\bea
\Gamma_\kappa  &=& i  \frac{\rho z^2}{\rho^2+z^2}\Gamma_{11}\Big[ - \frac{1}{z} \Gamma_{12} + \frac{1}{\rho} (\sin\sigma \Gamma_1 - \cos\sigma \Gamma_2) \Gamma_6 
\nn\\
&& + ( \frac{\sqrt{\rho^2-z^2}}{z} \Gamma_3 + \Gamma_5)
(
\frac{1}{z} ( \cos\sigma \Gamma_1 + \sin\sigma \Gamma_2 ) - \frac{1}{\rho} \Gamma_6 
)
\Big] . 
\eea
The problem is then to check if there exists a set of projection conditions on constant
spinor $\epsilon_0$, upon which the supersymmetry condition $\Gamma_\kappa \epsilon
=\epsilon$ is satisfied at every point $(z,\sigma)$. Since the form of $\Gamma_\kappa$ is
quite involved, this is a nontrivial task. 

Our strategy is first to make an educated guess for (part of) the projection rule, which simplifies the form 
of
Killing spinor \eqref{ks}. Then we derive the rest of projection rule at a special point
on the worldsheet where $\Gamma_\kappa$ is simplified. In the end we will see that our
circular Wilson loop configuration is 1/6-BPS.

The hint for the projection rule is that we have chosen $5,6$ directions as the $S^2$ part of the active metric. This combined with the SUSY rule \eqref{proj} imply that on Killing spinor we should impose
\be
(\Gamma^{49}+\Gamma^{78})\epsilon_0 = (\Gamma^{56} + \hat\Gamma\Gamma_{11} ) \epsilon_0 = 0 . 
\ee
These are compatible with the 3/4-BPS condition \eqref{proj},  and one can easily check that the three of them together constitute 1/2-BPS condition $\epsilon_0$. With these conditions, we may rewrite \eqref{ks} as 
\bea
\epsilon &=& 
\left\{ 1 + \frac{r}{2} ( t \Gamma^{03}(1-\Gamma^{012}) - x \Gamma^{13}(1-\Gamma^{012}) - y \Gamma^{23}(1-\Gamma^{012}) ) 
\right\}
r^{\frac{\Gamma^{012}}{2}}
\nn\\
&\times &
(\cos\frac{\phi}{2} e^{\frac{\theta}{2} \hat{\Gamma}\Gamma_5} + \Gamma_{56} \sin\frac{\phi}{2}  e^{-\frac{\theta}{2} \hat{\Gamma}\Gamma_5} ) 
\epsilon_0 . 
\eea

To infer the rest of the projection rule, let us consider a specific point on worldvolume
first, e.g.  $z=\rho, \sigma=0$.  Then we have 
\bea
i\Gamma_\kappa \Gamma_{11}
&=&
 \Gamma_{12} + \Gamma_{26} + \Gamma_{15} + \Gamma_{56} 
.
\eea
For the Killing spinor, let us further set $t=x=y=0$ for simplicity. Then 
\be
\epsilon = \rho^{-\frac{\Gamma^{012}}{2}} \epsilon_0 . 
\ee
And it is easy to calculate that $\Gamma_\kappa\epsilon = \epsilon$ leads to 
\bea
\epsilon_0 &=& \rho^{+\frac{\Gamma^{012}}{2}} 
\frac{i \Gamma_{11}}{2} ( \Gamma_{12} + \Gamma_{26} + \Gamma_{15} + \Gamma_{56} ) 
\rho^{-\frac{\Gamma^{012}}{2}} \epsilon_0
\nn\\
&=& 
\frac{i \Gamma_{11}}{2} ( (\Gamma_{12} + \Gamma_{56}) \rho^{-\Gamma^{012}}  + ( \Gamma_{26} + \Gamma_{15} )) \ep_0 . 
\eea
In order to satisfy this for arbitrary $\rho$ it is then required 
\be 
\Gamma_{1256} \ep_0 =  \ep_0 . 
\ee
And when we combine this condition with the background projection rule \eqref{proj}, it is clear that we should 
also impose 
\be
\ep_0 = i \Gamma_{11} \Gamma_{15} \ep_0 . 
\ee
The projection rules we have identified so far leave 4 linearly independent spinors, thus make our 
solution 1/6-BPS of the background. We have also checked that for such Killing spinors the $\Gamma_{\kappa}$
projection is satisfied at every point of the worldvolume, by explicitly evaluating $(\Gamma_{\kappa}-1)\epsilon$ using a concrete basis for gamma matrices. One possible loophole of our argument
so far is that $\Gamma_{4789}\epsilon_0=\epsilon_0$ is rather ad hoc. However we have also verified
that it is impossible to satisfy the $\Gamma_\kappa$ projection everywhere on the worldsheet if we consider
the other half of the spinors $\Gamma_{4789}\epsilon_0=-\epsilon_0$. 

\section{Discussions}
\label{s6}
We have studied supersymmetric Wilson loops in ABJM model, in particular the ones with generic contours in this paper.
Their existence and the conserved supersymmetries are nontrivial on both sides of the AdS/CFT correspondence.
We have given explicit construction of Wilson loops both in ABJM model and the dual string theory, and checked the
supersymmetry and their expectation values. 
We believe our analysis in this paper provides another nontrivial evidence for the ABJM conjecture. 

Being $(N+M)\times(N+M)$, the supermatrix in \eqref{sm} can be considered as in fundamental representation 
of $SU(N|M)$. For the case of $\cN=4$ super Yang-Mills and IIB string theory in $AdS_5\times S^5$, it is known that Wilson loops
in generic tensor representations are dual to D3 or D5 branes \cite{Yamaguchi:2006tq}\cite{Gomis:2006sb}. It is natural to expect that string-like D-branes in 
$AdS_4\times \mathbb{CP}^3$ described e.g. in \cite{Kluson:2008wn} are dual to Wilson loops in different representations of supergroup $SU(N|M)$. It must be interesting to construct such operators in ABJM model and to calculate them in perturbation theory. We plan to address these issues in forthcoming publications. 

\begin{acknowledgments}
We thank Jun-Bao Wu for useful comments.
NK gratefully acknowledges support from the Institute for Advanced Study, where part of this work was carried out.
This work is supported by the sabbatical leave petition program (2012) of Kyung Hee University, the National Research Foundation of Korea (NRF) funded by the Ministry of Education, Science and Technology (MEST) of Korea with grant No. 670522, 2010-0023121, 2012046278 and also through the Center for Quantum Spacetime (CQUeST) of Sogang University with grant No. 2005-0049409.
\end{acknowledgments}
\bibliography{wl}
\end{document}